\documentclass[aps,prb,reprint,groupedaddress,showpacs,amsmath,amssymb,twocolumn,floatfix]{revtex4-1}
\usepackage{graphicx}
\usepackage{amsmath, amsthm, amssymb}
\usepackage{latexsym}
\usepackage{amssymb}
\usepackage{bm}
\usepackage{dcolumn}
\usepackage{epstopdf}
\usepackage{color}

\newcommand{\et}{{\it et al.}} 
\newcommand{\he}{$^{3}$He}
\newcommand{\acos}{\cos^{-1}}

\begin{document}

\title{New Chiral Phases of Superfluid \he\ Stabilized by Anisotropic Silica Aerogel}

\author{J. Pollanen}
\author{J.I.A. Li}
\author{C.A. Collett}
\author{W.J. Gannon}
\author{W.P. Halperin}
\email[]{w-halperin@northwestern.edu}
\author{J.A. Sauls}
\affiliation{Department of Physics and Astronomy, Northwestern University, Evanston, Illinois 60208, USA}

\date{\today}

\maketitle
A rich variety of Fermi systems condense by forming bound pairs, including high temperature \cite{Kir.95} and heavy fermion \cite{Hef.96} superconductors, Sr$_{2}$RuO$_{4}$ \cite{Mac.05}, cold atomic gases \cite{Chi.04}, and superfluid \he\ \cite{Osh.72}. Some of these form exotic quantum states having non-zero orbital angular momentum.  We have discovered, in the case of \he, that anisotropic disorder, engineered from highly porous silica aerogel, stabilizes a chiral superfluid state that otherwise would not exist.  Additionally, we find that the chiral axis of this state can be uniquely oriented with the application of a magnetic field perpendicular to the aerogel anisotropy axis. At sufficiently low temperature we observe a sharp transition from a uniformly oriented chiral state to a disordered structure consistent with locally ordered domains, contrary to expectations for a superfluid glass phase \cite{Vol.08}.

Superconducting states with non-zero orbital angular momentum, $L\neq0$, are characterized by a competitive, but essential, relationship with magnetism, strong normal-state anisotropy, or both \cite{Kir.95,Hef.96,Mac.05,Osh.72}.  Moreover, these states are strongly suppressed by disorder, an important consideration for applications \cite{Sca.04} and a signature of their unconventional behavior \cite{Tsu.62, Dal.95, Mac.05}.  Although liquid \he\ in its normal phase is perfectly isotropic, it becomes a $p$-wave superfluid at low temperatures with non-zero orbital and spin angular momenta, $L=S=1$ \cite{Vol.90}. One of its two superfluid phases in zero magnetic field is anisotropic having chiral symmetry, where the handedness results from the orbital motion of the bound \he\ pairs about an axis $\vec{\ell}$.  This chiral superfluid, called the A-phase or axial state, is stable at high pressure near the normal-to-superfluid transition, Fig.~1a-c, while the majority of the phase diagram is the non-chiral B-phase, having isotropic physical properties.  The stability of the A-phase is attributed to strong-coupling effects arising from collisions between \he\ quasiparticles \cite{Vol.90}.  However, in the presence of isotropic disorder these strong-coupling effects are reduced and the stable chiral phase disappears \cite{Moo.10, Pol.11}, Fig.~1a.  Here we show that {\it anisotropic} disorder can reverse this process and stabilize an anisotropic phase over the entire phase diagram, Fig.~1c.
\begin{figure*}
\centerline{\includegraphics[height=0.4\textheight]{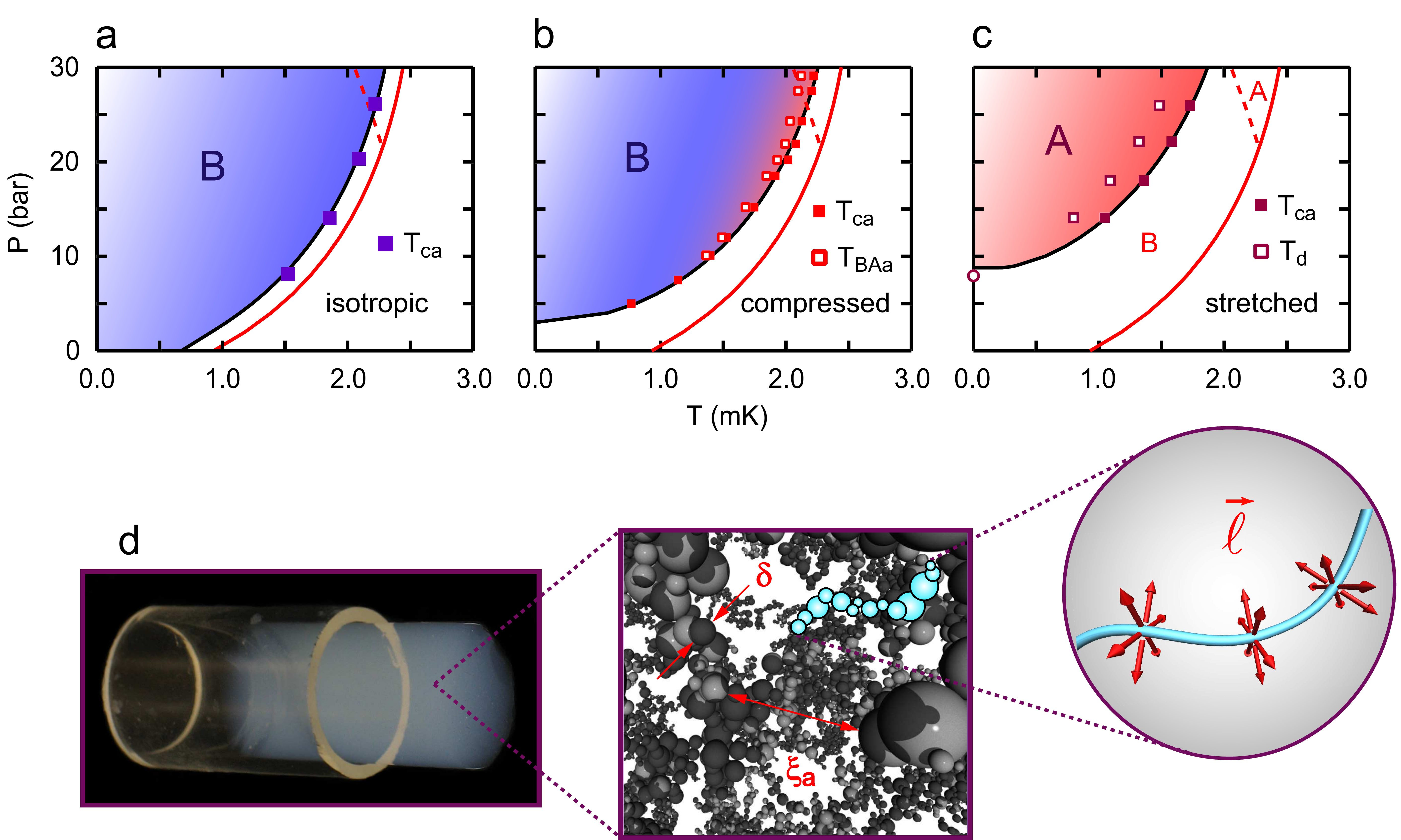}}
\caption{\label{fig1}{\bf{Superfluid phase diagrams.}} a-c) Comparison of phase diagrams of \he\ in silica aerogels with various types of anisotropy in the limit $H\rightarrow0$.  High spatial uniformity of the aerogels, better than 99.97\%, was established using optical cross polarization methods \cite{Pol.08} for a) and c). The solid black curves are fits to $T_{ca}(P)$ using the GL-scattering theory (Supplementary Information) and the solid (dashed) red curves correspond to the superfluid transition $T_{c}$ ($B \rightarrow A$ transition $T_{BA}$) for pure \he\ at $H=0$.  For subscripts, the lowercase $a$ signifies aerogel and uppercase $A$ and $B$ represent the corresponding phases of superfluid \he.  a) A uniformly isotropic 98.2\% porosity aerogel \cite{Pol.11}.  b) A 98.0\% porosity aerogel axially compressed by 10\% from torsional oscillator experiments by Bennett \et\ \cite{Ben.11}.  In both a) and b) the B-phase occupies the majority of the $P-T$ plane.  c) For the 97.5\% porosity stretched aerogel used in the present work the A-phase is stable throughout the phase diagram.  The open squares ($T_{d}$) mark the transition to a disordered chiral phase and the open circle at $T=0$ indicates a quantum critical point at which the superfluid order parameter vanishes (see Fig.~2b). d) Photograph of an aerogel with $\sim98$\% porosity, 1 cm in diameter, partially extracted from the glass tube in which it was grown.  The microstructure of aerogel based on numerical simulation of diffusion limited aggregation of silica particles, with diameter $\delta$, into a network of strands with most probable spacing, $\xi_{a}$.  The blow-up sketch shows the favorable directions (red arrows) of the chiral axes, $\vec{\ell}$, perpendicular to the strand.}
\end{figure*}

For many years it was thought to be impossible to introduce disorder into liquid \he\ since it is intrinsically chemically and isotopically pure at low temperatures.  Then it was discovered \cite{Por.95, Spr.95} that \he\ imbibed in $\sim98\%$ porosity silica aerogel, Fig.~1d, is a superfluid with a sharply defined  transition temperature \cite{Pol.11}, but reduced from that of pure \he.  To test predictions that isotropic disorder favors isotropic states and conversely \cite{Thu.98}, we have grown a 97.5\% porosity anisotropic aerogel with growth-induced radial compression \cite{Pol.08}, effectively stretching it along its cylinder axis by 14.3\%.  Experiments using uncharacterized stretched aerogels have been previously reported \cite{Elb.08, Dmi.10} and are in disagreement with the work presented here.  Silica aerogels, as in Fig.~1d, are formed  by silica particles $\approx 3$ nm in diameter,  precipitated from a tetramethylorthosilicate solution, and aggregated in a diffusion limited process.  After supercritical drying we obtain a cylinder as shown in Fig.~1d.  Our numerical simulation of this process indicates that the gel particles form strands, see Fig.~1d.  According to theory \cite{Rai.77}, $\vec{\ell}$ is constrained to point outward from the strand.  Stretching an aerogel tends to align the strands, thereby forcing the chiral axis, on average, to be in an easy-plane perpendicular to the aerogel anisotropy axis \cite{Vol.08}.  On the other hand, for an axially compressed aerogel $\vec{\ell}$ should be aligned with the compression axis \cite{Vol.08}.

We use pulsed nuclear magnetic resonance (NMR) (Methods Section), to identify the \he\ superfluid state and determine the magnitude of its order parameter, $\Delta$.   The equilibrium nuclear magnetization is tipped by a pulsed magnetic field to an angle $\beta$ away from the static applied magnetic field, $\vec{H}$, and the Fourier transform of its free precession is the NMR spectrum. The frequency shift of the spectrum, $\Delta\omega$, from the normal-state value centered at the Larmor frequency, $\omega_{L}$, is a direct measure of the temperature dependent order parameter, $\Delta(T)$.  For the A- and B-phases of \he\ \cite {Vol.90} the frequency shifts are,
\begin{eqnarray}
 \mbox{A-phase:}     \nonumber \\
   \Delta\omega_{A}(\beta)&=&\frac{\Omega^{2}_{A}}{2\omega_{L}}\left(\frac{1}{4}+\frac{3}{4}\cos(\beta)\right) \nonumber \\ \Omega^{2}_{A}&=&\Delta^{2}_{A}\left(\frac{12}{5} \,\,\frac{\gamma^2\lambda_{D}N_{F}}{\chi_{A}}\right),
\end{eqnarray}
\begin{eqnarray}
 \mbox{B-phase:}     \nonumber \\
   \Delta\omega_{B}(\beta)&=& 0 \,\,\,\,\,\,\,\,\,\,  \beta < 104^{\circ} \nonumber \\ 
   \Omega^{2}_{B} &=& \Delta^{2}_{B}\left(6 \,\,\frac{\gamma^2\lambda_{D}N_{F}}{\chi_B}\right),
\end{eqnarray}
\\
\noindent	
where $\Omega^2_{A,B}$ is the square of the longitudinal resonance frequency, which is proportional to $\Delta^2_{A,B}$ and determined by constants of the normal fluid: gyromagnetic ratio, $\gamma$; dimensionless dipole coupling constant, $\lambda_{D}$; single spin density of states at the Fermi level, $N_{F}$; as well as the magnetic susceptibilities $\chi_{A}$ and $\chi_B$.  The A-phase susceptibility is temperature independent and equal to the normal-state value, $\chi_{N}$; however, $\chi_{B}$ is temperature dependent and suppressed relative to $\chi_{N}$.   

For \he\ in our anisotropic aerogel we observe a sharp transition to a superfluid state at a temperature, $T_{ca}$, marked by the onset of NMR frequency shifts for small tip angle, $\beta = 8^\circ$, Fig.~2a.  Our comparison of the data with either of the two pure \he\ states based on Eq.~1 and 2, suggests that the superfluid is an axial $p$-wave state like the A-phase.  In the Ginzburg-Landau (GL) limit, $T\lesssim T_{c}$, the initial slope of $\Delta\omega_A$ is proportional to the square of the order parameter magnitude $\Delta^{2}_{A0}$ where $\Delta^2_A(T) = \Delta^2_{A0}(1-T/T_{c})$.  We find in our anisotropic aerogel that $\Delta^{2}_{A0}$ has a linear pressure dependence, as shown in Fig.~2b, just as is the case for pure \he-A \cite{Sch.93, Ran.96}.  It is, however, substantially reduced in magnitude and extrapolates to a critical pressure of $P_{c} = 7.9$ bar at $T=0$ where the order parameter vanishes at a quantum critical point \cite{Mat.97}.  We have analyzed the transition temperatures $T_{ca}(P)$ using GL theory \cite{Sau.03} (Supplementary Information) to determine  the mean free path $\lambda = 113$ nm and the silica particle correlation length $\xi_a= 39$ nm.  Our calculated phase diagram compares very well with the data in Fig.~1c.  With the same parameters we have also calculated $\Delta^{2}_{A0}$  (Supplementary Information), shown in Fig.~2b, and find excellent agreement with the order parameter determined from the frequency shifts, Eq.~1, consistent with a suppressed A-phase.  Moreover, extrapolation of  $\Delta\omega(P) \rightarrow 0$ is at a pressure that coincides with the critical pressure from the calculated phase diagram, demonstrating  important consistency between experiment and  theory.
\begin{figure*}
\centerline{\includegraphics[height=0.45\textheight]{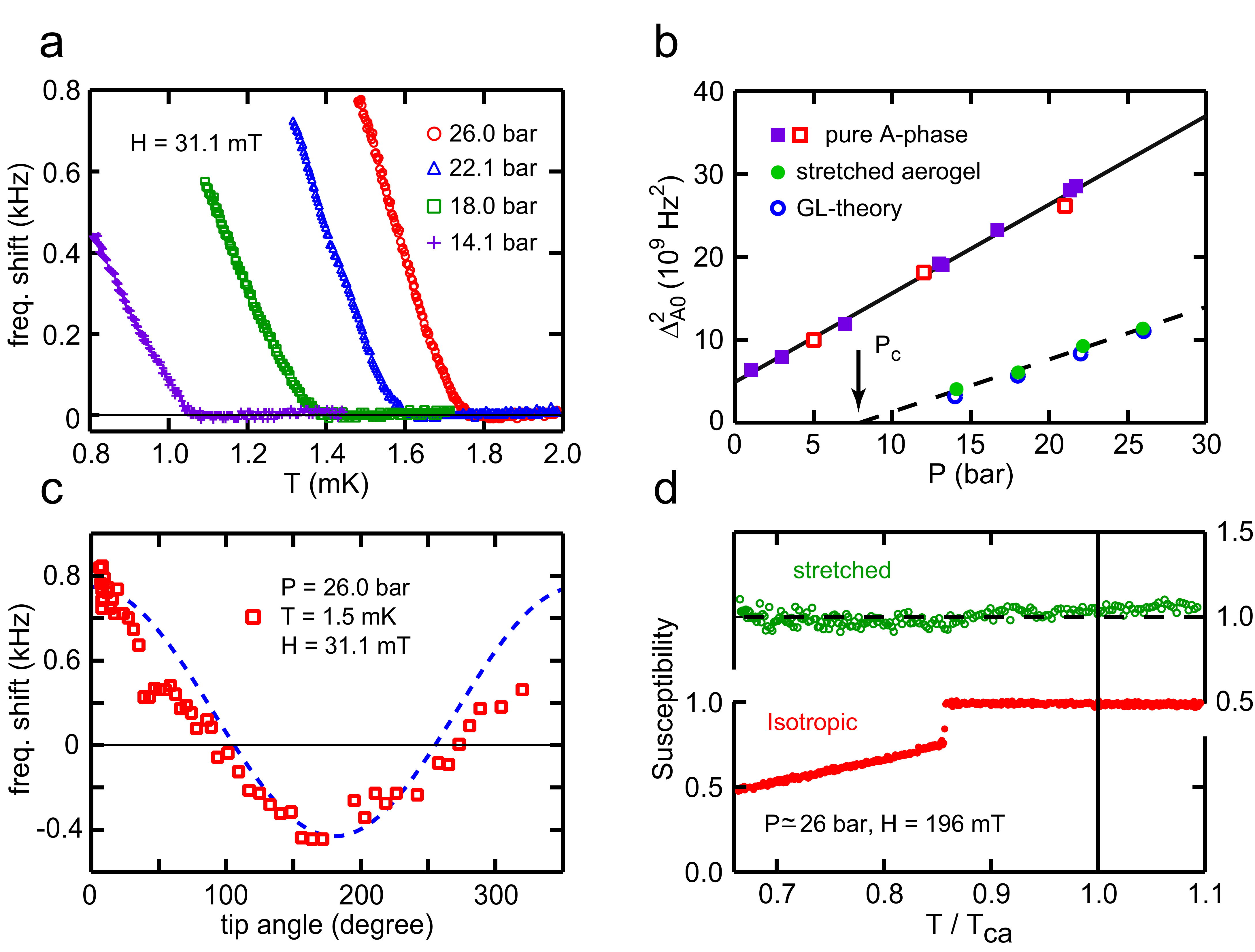}}
\caption{\label{fig2}{\bf{Identification of the superfluid state.}}  
a) The frequency shift of the NMR spectrum for a small tip angle, $\beta=8^{\circ}$, as determined from gaussian fits.  The superfluid transition temperature at different pressures is marked by the onset of positive frequency shifts.  The NMR spectra have been corrected for fast exchange of  liquid spins with several layers of paramagnetic solid \he\ on the aerogel surface \cite{Spr.95, Col.09}.
b) The square of the order parameter magnitude as a function of pressure in our aerogel (closed green circles) was calculated from the initial slopes in a) using Eq. 1, showing the suppression relative to pure \he-A (open and closed squares) \cite{Sch.93, Ran.96}.  The experimental results are compared to the predictions of the GL-theory (open blue circles) (Supplementary Information).  The critical pressure, $P_{c}=7.9$ bar is determined from an extrapolation (dashed line) of $\Delta^{2}_{A0}$ to zero, indicated by the arrow.  The solid line is a linear fit to the pure A-phase data.
c) Identification of the superfluid as the axial $p$-wave state follows from the agreement with theory \cite{Bri.75} (dashed blue curve) for the dependence of $\Delta\omega$ on tip angle, $T_{d} < T < T_{c}$.  
d) The susceptibility for stretched and isotropic aerogel samples taken on warming at a pressure $P\simeq26$ bar in $H = 196$ mT after subtraction of a paramagnetic background from solid \he\ \cite{Spr.95}.  For the isotropic aerogel the field-induced A-phase region has a constant susceptibility below $T_{ca}$ and a very sharp jump at the BA-transition \cite{Pol.11}.  A similar jump in susceptibility would be anticipated for the stretched aerogel if the \he\ were in the B-phase or any other non-ESP phase.}
\end{figure*}

The tip angle dependence of the NMR frequency shift is a fingerprint of a specific $p$-wave state \cite{Vol.90}.  Our measurements of $\Delta\omega(\beta)$, Fig.~2c, follow the expected behavior of the chiral axial state \cite{Bri.75}, Eq.~1.  The magnitude of the shift is reduced compared to that of pure \he-A as shown in Fig.~2b.  Furthermore, the axial state belongs to the class of equal spin pairing (ESP) states, having the same susceptibility as the normal Fermi liquid.  We have directly demonstrated that this is the case as shown in Fig.~2d.   

Lastly, we have looked for a possible phase transition to some other $p$-wave state by cooling to our lowest temperatures, $\lesssim 650\,\mu$K, in a substantial magnetic field of 196 mT, and measuring $\chi$ and $\Delta\omega$ on warming.  For example, if there had been a transition to a B-phase on cooling we would have observed a discontinuous increase of $\chi$ on warming from the B-phase to a magnetic field-induced A-phase, as is characteristic of the first order B-to-A transition in an isotropic aerogel (Fig.~2d) \cite{Pol.11}.  Correspondingly, we would also have observed a discontinuous change in $\Delta\omega$.  The absence of these discontinuities demonstrates that an ESP superfluid state in stretched aerogel is stable down to our lowest temperatures.

Our three independent NMR measurements  show that anisotropic stretched aerogel stabilizes anisotropic superfluidity throughout the phase diagram, Fig.~1c.  This is in stark contrast to \he\ in the presence of isotropic disorder \cite{Moo.10, Pol.11}, Fig.~1a, or for anisotropic disorder established by axial compression, Fig.~1b \cite{Ben.11}.  In both of these latter cases the B-phase is dominant.  Understanding the stability of chiral superfluidity, stabilized by anisotropy, is an interesting open problem, in part because topologically non-trivial pairing states have unusual structure.

Our NMR results reported in Fig.~2a-c were restricted to temperatures within $\sim 20$\% of $T_{ca}$.  In this high temperature region there is negligible increase in the NMR linewidth as compared to the normal fluid, see Fig.~3, indicating that the chiral axis, $\vec{\ell}$, has a well-defined direction orthogonal to both the aerogel anisotropy axis \cite{Vol.08} and the magnetic field.  This orientation minimizes the magnetic dipole energy producing the so-called dipole-locked condition on which the validity of Eq.~1 depends \cite{Vol.90}, and with which our tip angle results agree.  The combined effects of the dipole energy and aerogel anisotropy  define a uniform direction of $\vec{\ell}$ for the A-phase of superfluid \he\ in our stretched aerogel.
\begin{figure*}
\centerline{\includegraphics[height=0.35\textheight]{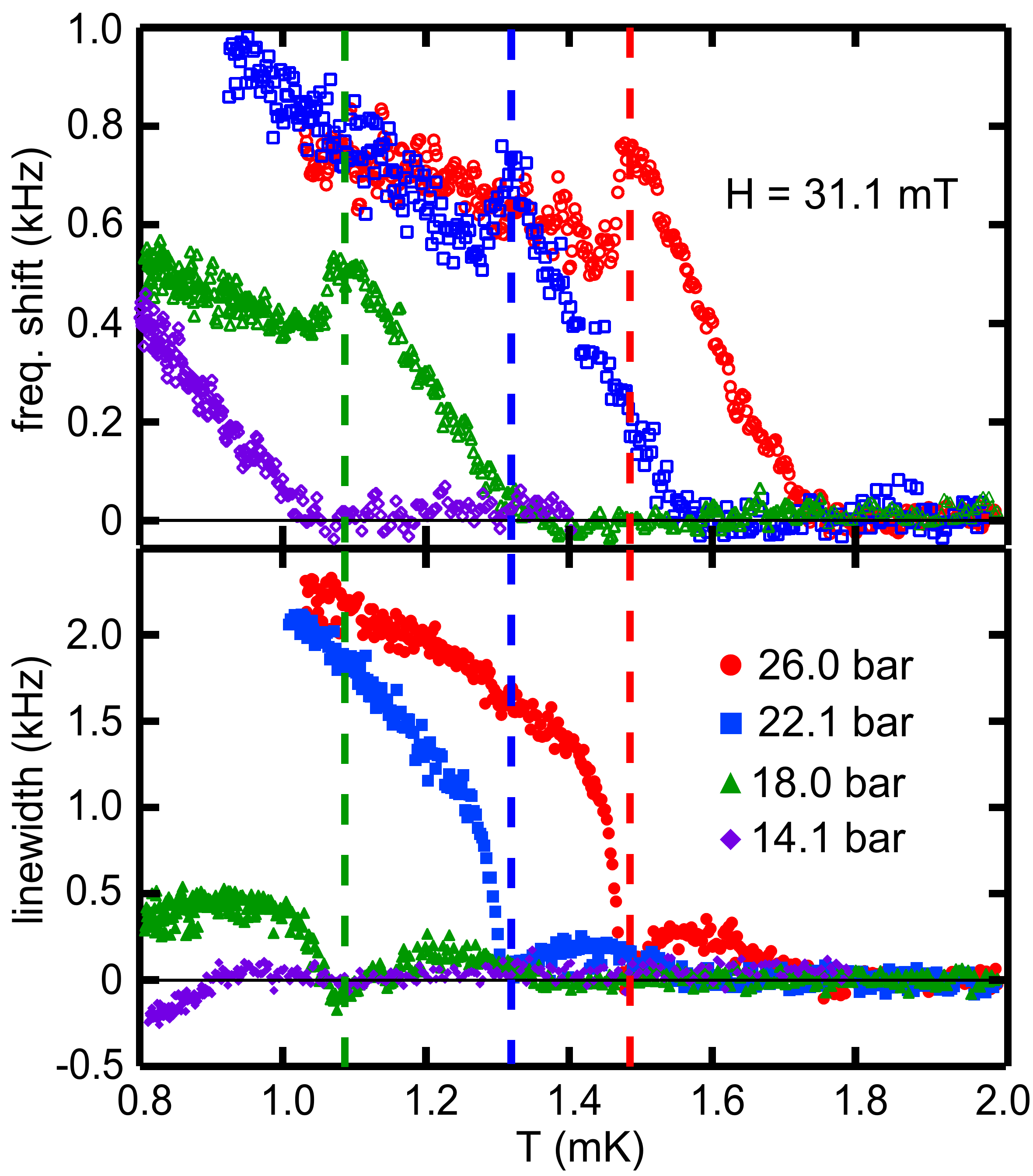}}
\caption{\label{fig3}{\bf{The disorder transition.}} Below the disorder transition, $T<T_{d}$ (vertical dashed lines), the NMR frequency shift drops and the linewidth abruptly increases.  The frequency shift is defined as the first moment of the NMR spectrum and the linewidth is the full width at half maximum.  Both quantities have been corrected for the effect of fast exchange between liquid and surface solid \he\ spins \cite{Spr.95, Col.09}. Data were taken on warming.}
\end{figure*}

At lower temperatures, we have discovered a different phase characterized by a significantly broadened NMR line.  The data indicate that this superfluid state is no longer the homogeneous A-phase with $\vec{\ell} \perp \vec{H}$.  On warming there is an abrupt increase in the first moment of the NMR spectrum (frequency shift) and a corresponding decrease in the linewidth at a well-defined temperature, $T_d$ (Fig.~3).  The transition at $T_d$ is independent of magnetic field for $31.1<H<196$ mT.  In addition, the magnetic susceptibility does not change at this temperature; however, $T_{d}$ is pressure dependent as shown by the open squares in the phase diagram of Fig.~1c.  We interpret these results as evidence for a disordered ESP phase with a possible phase transition at $T_{d}$.    

The frequency shift is a measure of $\Delta_A$, given in Eq.~1 for the dipole-locked case where the angle between the chiral axis and the magnetic field is $\theta=\acos(\vec{\ell} \cdot \vec{H})=90^\circ$.  However, in general the shift also depends on $\theta$ which for small tip angles $\Delta\omega(\theta)$ reduces to \cite{Bun.93},
	\begin{equation}
	\Delta\omega(\theta)=-\frac{\Omega^{2}_{A}}{2\omega_{L}}\cos(2\theta)\label{3}.
	\end{equation} 
\noindent
This allows us to extract the angular distribution, $P(\theta)$, of $\vec{\ell}$ relative to $\vec{H}$, from the NMR spectra both above and below $T_{d}$. The observed spectrum is given by the convolution product of the normal-state lineshape and the frequency distribution $P(\omega)$ that corresponds to $P(\theta)$ determined by Eq.~3 (Supplementary Information).  Above $T_{d}$ the NMR spectrum (Fig.~4a) can be fit with an $\vec{\ell}$-distribution (Fig.~4b) that has a single component centered at $\theta = 90^\circ$, as sketched in Fig.~4c, with angular spread  $\Delta\theta=13^{\circ} \pm 2.3^{\circ}$.  Examination of the spectrum below $T_{d}$, Fig.~4a, indicates that it is bimodal, suggesting that $P(\theta)$ has three main components, which we take to be gaussian functions with adjustable position, width, and relative weight.  From our fits we determine these parameters and obtain $P(\theta)$, Fig.~4b.  For the spectra below $T_{d}$, the disordered state is composed of three domains.  Approximately $1/3$ of the sample remains with $\vec{\ell} \perp \vec{H}$ while the majority, $\sim2/3$ of the distribution, has nearly zero frequency shift with $\theta=44 \pm 0.5^\circ$.  The theory, Eq.~3, does not distinguish between angles that are symmetric about $\theta=90^\circ$, and thus we assume that $P(\theta)=P(180^{\circ}-\theta)$ as shown in Fig.~4d.
\begin{figure*}
\centerline{\includegraphics[height=0.6\textheight]{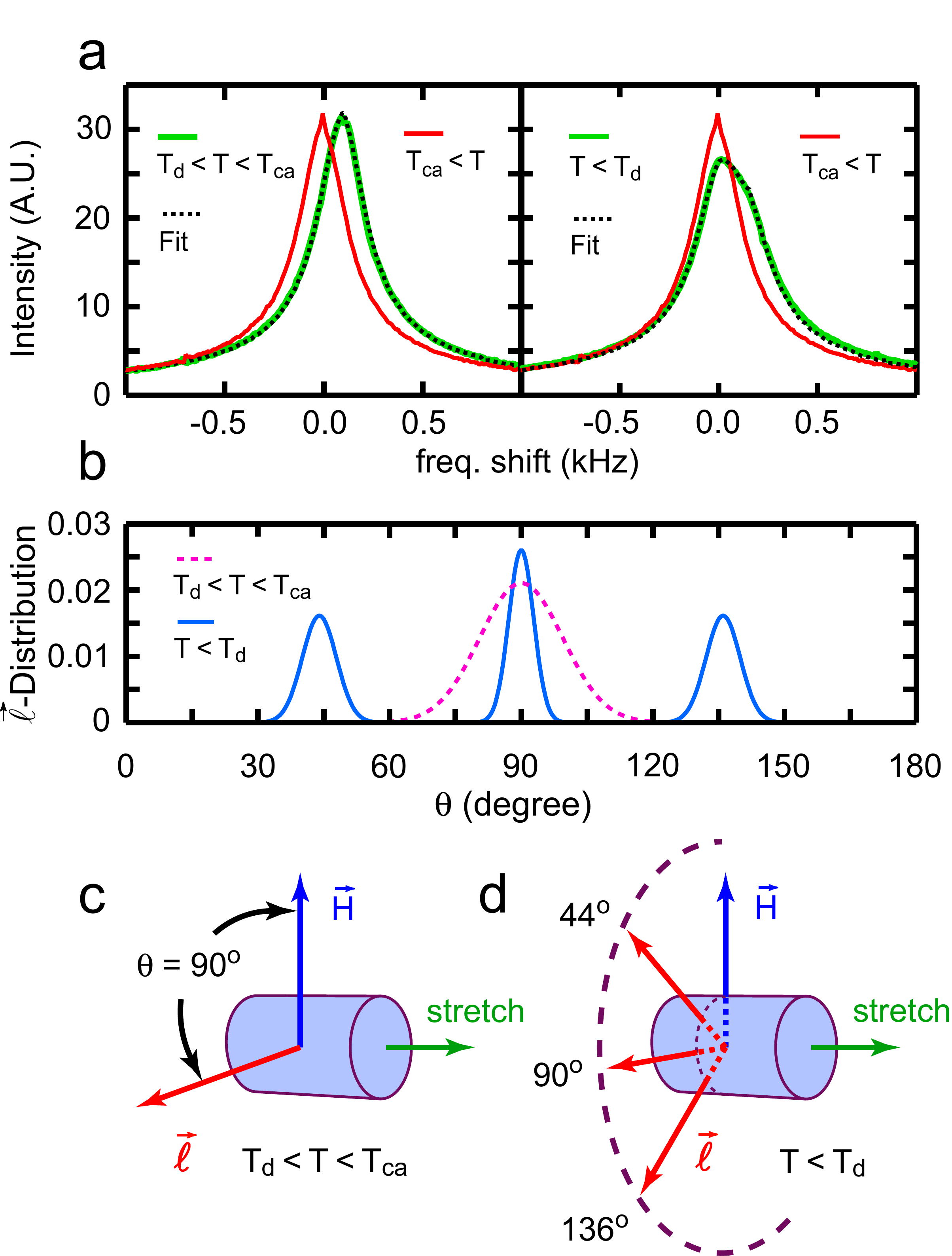}}
\caption{\label{fig4}{\bf{Distribution of the direction of the chiral axis.}}  
a) The convolution product of the normal-state NMR line (red curve) with a best-fit angular distribution of the chiral axis, $P(\theta)$, gives an excellent representation (dashed black curves) of the NMR spectra for the superfluid state above (left panel, bold green curve) and below (right panel, bold green curve) the disorder transition (Supplementary Information).  The spectra were obtained with $H=31.1$ mT.
b) The angular distributions for the chiral axis, $\vec{\ell}$, determined from a). The dashed pink curve is the distribution for $T_d<T<T_{ca}$.  
c) Stretching the aerogel along the cylinder axis forces $\vec{\ell}$ into an easy-plane perpendicular to the direction of strain \cite{Vol.08}.  To minimize the dipole energy $\vec{\ell}$ must be perpendicular to $\vec{H}$, as illustrated.  
d) According to our analysis the distribution splits into three $\vec{\ell}$-domains for $T<T_{d}$.}
\end{figure*}
Our interpretation of the disordered state in terms of distinct $\vec{\ell}$-domains is different from the two-dimensional ``orbital glass'' phase defined by a random distribution of $\vec{\ell}$ confined to a plane perpendicular to the stretching axis \cite{Vol.08}.  It is noteworthy that the possible existence of orbital domains, such as we observe here, has been central to the discussion of time-reversal symmetry breaking in the unconventional superconducting state in Sr$_{2}$RuO$_{4}$ \cite{Fer.11}.

\begin{acknowledgments}
We are grateful to J.M. Parpia, V.V. Dmitriev, G.E. Volovik, N. Mulders, K.R. Shirer, A.M. Mounce and Y. Lee for helpful discussion and for support from the National Science Foundation, DMR-1103625, DMR-0805277, and DMR-1106315.
\end{acknowledgments}

\section{Methods}
In our experiments the external magnetic field $\vec{H}$ was 31.1, 95.5, or 196 mT and oriented perpendicular to the aerogel cylinder axis.  The sample was 4.93 mm long and had a diameter of 3.43 mm.  The susceptibility was obtained by numerically integrating the phase-corrected absorption spectrum, while the frequency shift and linewidth were determined from the power spectrum.  The sample was cooled using adiabatic nuclear demagnetization of PrNi$_{5}$.  NMR on $^{195}$Pt was used for thermometry at $H=95.5$ and 196 mT and calibrated relative to the known phase diagram of pure superfluid \he\ from a volume of liquid outside of the aerogel sample equaling 30\% of the total liquid.  For data acquired at $H=31.1$ mT, temperature was determined from the Curie-Weiss dependence of the solid \he\ adsorbed to the aerogel surface \cite{Spr.95, Col.09}.
\clearpage
\bibliography{NPPollRef}

\begin{thebibliography}{10}
\expandafter\ifx\csname url\endcsname\relax
  \def\url#1{\texttt{#1}}\fi
\expandafter\ifx\csname urlprefix\endcsname\relax\def\urlprefix{URL }\fi
\providecommand{\bibinfo}[2]{#2}
\providecommand{\eprint}[2][]{\url{#2}}

\bibitem{Kir.95}
\bibinfo{author}{Kirtley, J.~R.} \emph{et~al.}
\newblock \bibinfo{title}{Symmetry of the order parameter in the high-{T}$_{c}$
  superconductor {Y}{B}a$_{2}${C}u$_{3}${O}$_{7-\delta}$}.
\newblock \emph{\bibinfo{journal}{Nature}} \textbf{\bibinfo{volume}{373}},
  \bibinfo{pages}{225--228} (\bibinfo{year}{1995}).

\bibitem{Hef.96}
\bibinfo{author}{Heffner, R.~H.} \& \bibinfo{author}{Norman, M.~R.}
\newblock \bibinfo{title}{Heavy fermion superconductivity}.
\newblock \emph{\bibinfo{journal}{Comments Cond. Matter Phys.}}
  \textbf{\bibinfo{volume}{17}}, \bibinfo{pages}{361--408}
  (\bibinfo{year}{1996}).

\bibitem{Mac.05}
\bibinfo{author}{Mackenzie, A.~P.} \& \bibinfo{author}{Maeno, Y.}
\newblock \bibinfo{title}{The superconductivity of {S}r$_{2}${R}u{O}$_{4}$ and
  the physics of spin-triplet pairing}.
\newblock \emph{\bibinfo{journal}{Rev. Mod. Phys.}}
  \textbf{\bibinfo{volume}{75}}, \bibinfo{pages}{657--712}
  (\bibinfo{year}{2005}).

\bibitem{Chi.04}
\bibinfo{author}{Chin, C.} \emph{et~al.}
\newblock \bibinfo{title}{Observation of the pairing gap in a strongly
  interacting fermi gas}.
\newblock \emph{\bibinfo{journal}{Science}} \textbf{\bibinfo{volume}{305}},
  \bibinfo{pages}{1128--1130} (\bibinfo{year}{2004}).

\bibitem{Osh.72}
\bibinfo{author}{Osheroff, D.~D.}, \bibinfo{author}{Richardson, R.~C.} \&
  \bibinfo{author}{Lee, D.~M.}
\newblock \bibinfo{title}{Evidence for a new phase of solid {H}e$^{3}$}.
\newblock \emph{\bibinfo{journal}{Phys. Rev. Lett.}}
  \textbf{\bibinfo{volume}{28}}, \bibinfo{pages}{885--888}
  (\bibinfo{year}{1972}).

\bibitem{Vol.08}
\bibinfo{author}{Volovik, G.~E.}
\newblock \bibinfo{title}{On {L}arkin-{I}mry-{M}a state of $^{3}${H}e-{A} in
  aerogel}.
\newblock \emph{\bibinfo{journal}{J. Low Temp. Phys.}}
  \textbf{\bibinfo{volume}{150}}, \bibinfo{pages}{453--463}
  (\bibinfo{year}{2008}).

\bibitem{Sca.04}
\bibinfo{author}{Scanlan, R.~M.}, \bibinfo{author}{Malozemoff, A.~P.} \&
  \bibinfo{author}{Larbalestier, D.~C.}
\newblock \bibinfo{title}{Superconducting materials for large scale
  applications}.
\newblock \emph{\bibinfo{journal}{Proc. IEEE}} \textbf{\bibinfo{volume}{92}},
  \bibinfo{pages}{1639--1654} (\bibinfo{year}{2004}).

\bibitem{Tsu.62}
\bibinfo{author}{Tsuneto, T.}
\newblock \bibinfo{title}{On dirty superconductors}.
\newblock \emph{\bibinfo{journal}{Inst. Solid State Phys., Univ. Tokyo Tech.
  Rep.}} \textbf{\bibinfo{volume}{47}}, \bibinfo{pages}{Ser. A}
  (\bibinfo{year}{1962}).

\bibitem{Dal.95}
\bibinfo{author}{Dalichaouch, Y.} \emph{et~al.}
\newblock \bibinfo{title}{Impurity scattering and triplet superconductivity in
  {U}{P}t$_{3}$}.
\newblock \emph{\bibinfo{journal}{Phys. Rev. Lett.}}
  \textbf{\bibinfo{volume}{75}}, \bibinfo{pages}{3938--3941}
  (\bibinfo{year}{1995}).

\bibitem{Vol.90}
\bibinfo{author}{Vollhardt, D.} \& \bibinfo{author}{W{\"o}lfle, P.}
\newblock \emph{\bibinfo{title}{The Superfluid Phases of Helium 3}}
  (\bibinfo{publisher}{Taylor and Francis}, \bibinfo{year}{1990}).

\bibitem{Moo.10}
\bibinfo{author}{Moon, B.~H.} \emph{et~al.}
\newblock \bibinfo{title}{Ultrasound attenuation and a {P}-{B}-{T} phase
  diagram of superfluid $^{3}\text{H}\text{e}$ in 98\% aerogel}.
\newblock \emph{\bibinfo{journal}{Phys. Rev. B}} \textbf{\bibinfo{volume}{81}},
  \bibinfo{pages}{134526} (\bibinfo{year}{2010}).

\bibitem{Pol.11}
\bibinfo{author}{Pollanen, J.}, \bibinfo{author}{Li, J. I.~A.},
  \bibinfo{author}{Collett, C.~A.}, \bibinfo{author}{Gannon, W.~J.} \&
  \bibinfo{author}{Halperin, W.~P.}
\newblock \bibinfo{title}{Identification of superfluid phases of $^{3}${H}e in
  uniformly isotropic 98.2\% aerogel}.
\newblock \emph{\bibinfo{journal}{Phys. Rev. Lett.}}
  \textbf{\bibinfo{volume}{107}}, \bibinfo{pages}{195301}
  (\bibinfo{year}{2011}).

\bibitem{Pol.08}
\bibinfo{author}{Pollanen, J.} \emph{et~al.}
\newblock \bibinfo{title}{Globally anisotropic high porosity silica aerogels}.
\newblock \emph{\bibinfo{journal}{J. Non-Crystalline Solids}}
  \textbf{\bibinfo{volume}{354}}, \bibinfo{pages}{4668} (\bibinfo{year}{2008}).

\bibitem{Ben.11}
\bibinfo{author}{Bennett, R.~G.} \emph{et~al.}
\newblock \bibinfo{title}{Modification of the $^3${H}e phase diagram by
  anisotropic disorder}.
\newblock \emph{\bibinfo{journal}{Phys. Rev. Lett.}}
  \textbf{\bibinfo{volume}{107}}, \bibinfo{pages}{235504}
  (\bibinfo{year}{2011}).

\bibitem{Por.95}
\bibinfo{author}{Porto, J.~V.} \& \bibinfo{author}{Parpia, J.~M.}
\newblock \bibinfo{title}{Superfluid $^{3}${H}e in aerogel}.
\newblock \emph{\bibinfo{journal}{Phys. Rev. Lett.}}
  \textbf{\bibinfo{volume}{74}}, \bibinfo{pages}{4667} (\bibinfo{year}{1995}).

\bibitem{Spr.95}
\bibinfo{author}{Sprague, D.~T.} \emph{et~al.}
\newblock \bibinfo{title}{Homogeneous equal-spin pairing superfluid state of
  $^{3}${H}e in aerogel}.
\newblock \emph{\bibinfo{journal}{Phys. Rev. Lett.}}
  \textbf{\bibinfo{volume}{75}}, \bibinfo{pages}{661} (\bibinfo{year}{1995}).

\bibitem{Thu.98}
\bibinfo{author}{Thuneberg, E.~V.}, \bibinfo{author}{Yip, S.~K.},
  \bibinfo{author}{Fogelstr{\"o}m, M.} \& \bibinfo{author}{Sauls, J.~A.}
\newblock \bibinfo{title}{Models for superfluid $^{3}${H}e in aerogel}.
\newblock \emph{\bibinfo{journal}{Phys. Rev. Lett.}}
  \textbf{\bibinfo{volume}{80}}, \bibinfo{pages}{2861} (\bibinfo{year}{1998}).

\bibitem{Elb.08}
\bibinfo{author}{Elbs, J.}, \bibinfo{author}{Bunkov, Y.~M.},
  \bibinfo{author}{Collin, E.} \& \bibinfo{author}{Godfrin, H.}
\newblock \bibinfo{title}{Strong orientational effect of stretched aerogel on
  the $^{3}${H}e order parameter}.
\newblock \emph{\bibinfo{journal}{Phys. Rev. Lett.}}
  \textbf{\bibinfo{volume}{100}}, \bibinfo{pages}{215304}
  (\bibinfo{year}{2008}).

\bibitem{Dmi.10}
\bibinfo{author}{Dmitriev, V.~V.} \emph{et~al.}
\newblock \bibinfo{title}{Orbital glass and spin glass states of $^{3}${H}e-{A}
  in aerogel}.
\newblock \emph{\bibinfo{journal}{JETP Lett.}} \textbf{\bibinfo{volume}{91}},
  \bibinfo{pages}{599--606} (\bibinfo{year}{2010}).

\bibitem{Rai.77}
\bibinfo{author}{Rainer, D.} \& \bibinfo{author}{Vuorio, M.}
\newblock \bibinfo{title}{Small objects in supefluid $^{3}${H}e}.
\newblock \emph{\bibinfo{journal}{J. Phys. C: Solid State Phys.}}
  \textbf{\bibinfo{volume}{10}}, \bibinfo{pages}{3093} (\bibinfo{year}{1977}).

\bibitem{Sch.93}
\bibinfo{author}{Schiffer, P.~E.}
\newblock \emph{\bibinfo{title}{Studies of the Superfluid Phases of Helium
  Three and The Magnetization of Thin Solid Films of Helium Three}}.
\newblock Ph.D. thesis, \bibinfo{school}{Stanford University}
  (\bibinfo{year}{1993}).

\bibitem{Ran.96}
\bibinfo{author}{Rand, M.~R.}
\newblock \emph{\bibinfo{title}{Nonlinear Spin Dynamics and Magnetic Field
  Distortion of the Superfluid $^{3}${H}e-{B} Order Parameter}}.
\newblock Ph.D. thesis, \bibinfo{school}{Northwestern University}
  (\bibinfo{year}{1996}).

\bibitem{Mat.97}
\bibinfo{author}{Matsumoto, K.} \emph{et~al.}
\newblock \bibinfo{title}{Quantum phase transition of $^{3}${H}e in aerogel at
  a nonzero pressure}.
\newblock \emph{\bibinfo{journal}{Phys. Rev. Lett.}}
  \textbf{\bibinfo{volume}{79}}, \bibinfo{pages}{253--256}
  (\bibinfo{year}{1997}).

\bibitem{Sau.03}
\bibinfo{author}{Sauls, J.~A.} \& \bibinfo{author}{Sharma, P.}
\newblock \bibinfo{title}{Impurity effects on the {A}$_{1}$-{A}$_{2}$ splitting
  of superfluid $^{3}${H}e in aerogel}.
\newblock \emph{\bibinfo{journal}{Phys. Rev. B}} \textbf{\bibinfo{volume}{68}},
  \bibinfo{pages}{224502} (\bibinfo{year}{2003}).

\bibitem{Col.09}
\bibinfo{author}{Collin, E.}, \bibinfo{author}{Triqueneaux, S.},
  \bibinfo{author}{Bunkov, Y.~M.} \& \bibinfo{author}{Godfrin, H.}
\newblock \bibinfo{title}{Fast-exchange model visualized with $^{3}${H}e
  confined in aerogel: {A} {F}ermi liquid in contact with a ferromagnetic
  solid}.
\newblock \emph{\bibinfo{journal}{Phys. Rev. B}} \textbf{\bibinfo{volume}{80}},
  \bibinfo{pages}{094422} (\bibinfo{year}{2009}).

\bibitem{Bri.75}
\bibinfo{author}{Brinkman, W.~F.} \& \bibinfo{author}{Smith, H.}
\newblock \bibinfo{title}{Frequency shifts in pulsed {NMR} for
  $^{3}${H}e({A})}.
\newblock \emph{\bibinfo{journal}{Phys. Lett.}} \textbf{\bibinfo{volume}{51}},
  \bibinfo{pages}{449--450} (\bibinfo{year}{1975}).

\bibitem{Bun.93}
\bibinfo{author}{Bunkov, Y.~M.} \& \bibinfo{author}{Volovik, G.~E.}
\newblock \bibinfo{title}{On the possibility of the homogeneously precessing
  domain in bulk $^{3}${H}e-{A}}.
\newblock \emph{\bibinfo{journal}{Europhys. Lett.}}
  \textbf{\bibinfo{volume}{21}}, \bibinfo{pages}{837--843}
  (\bibinfo{year}{1993}).

\bibitem{Fer.11}
\bibinfo{author}{Ferguson, D.~G.} \& \bibinfo{author}{Goldbart, P.~M.}
\newblock \bibinfo{title}{Penetration of nonintegral magnetic flux through a
  domain-wall bend in time-reversal symmetry broken superconductors}.
\newblock \emph{\bibinfo{journal}{Phys. Rev. B}} \textbf{\bibinfo{volume}{84}},
  \bibinfo{pages}{014523} (\bibinfo{year}{2011}).

\end{thebibliography}


\begin{thebibliography}{1}
\expandafter\ifx\csname url\endcsname\relax
  \def\url#1{\texttt{#1}}\fi
\expandafter\ifx\csname urlprefix\endcsname\relax\def\urlprefix{URL }\fi
\providecommand{\bibinfo}[2]{#2}
\providecommand{\eprint}[2][]{\url{#2}}

\bibitem{Thu.98}
\bibinfo{author}{Thuneberg, E.~V.}, \bibinfo{author}{Yip, S.~K.},
  \bibinfo{author}{Fogelstr{\"o}m, M.} \& \bibinfo{author}{Sauls, J.~A.}
\newblock \bibinfo{title}{Models for superfluid $^{3}${H}e in aerogel}.
\newblock \emph{\bibinfo{journal}{Phys. Rev. Lett.}}
  \textbf{\bibinfo{volume}{80}}, \bibinfo{pages}{2861} (\bibinfo{year}{1998}).

\bibitem{Sau.03}
\bibinfo{author}{Sauls, J.~A.} \& \bibinfo{author}{Sharma, P.}
\newblock \bibinfo{title}{Impurity effects on the {A}$_{1}$-{A}$_{2}$ splitting
  of superfluid $^{3}${H}e in aerogel}.
\newblock \emph{\bibinfo{journal}{Phys. Rev. B}} \textbf{\bibinfo{volume}{68}},
  \bibinfo{pages}{224502} (\bibinfo{year}{2003}).

\bibitem{Note1}
\bibinfo{note}{Our definition of $x$ is a factor of 2 larger than that in
  Ref.~\cite {Thu.98}.}

\bibitem{Hal.90}
\bibinfo{author}{Halperin, W.~P.} \& \bibinfo{author}{Varoquax, E.}
\newblock \emph{\bibinfo{title}{Helium Three}} (\bibinfo{publisher}{Elsevier},
  \bibinfo{year}{1990}).

\bibitem{Sch.93}
\bibinfo{author}{Schiffer, P.~E.}
\newblock \emph{\bibinfo{title}{Studies of the Superfluid Phases of Helium
  Three and The Magnetization of Thin Solid Films of Helium Three}}.
\newblock Ph.D. thesis, \bibinfo{school}{Stanford University}
  (\bibinfo{year}{1993}).

\bibitem{Ran.96}
\bibinfo{author}{Rand, M.~R.}
\newblock \emph{\bibinfo{title}{Nonlinear Spin Dynamics and Magnetic Field
  Distortion of the Superfluid $^{3}${H}e-{B} Order Parameter}}.
\newblock Ph.D. thesis, \bibinfo{school}{Northwestern University}
  (\bibinfo{year}{1996}).

\bibitem{Cho.07}
\bibinfo{author}{Choi, H.}, \bibinfo{author}{Davis, J.~P.},
  \bibinfo{author}{Pollanen, J.}, \bibinfo{author}{Haard, T.~M.} \&
  \bibinfo{author}{Halperin, W.~P.}
\newblock \bibinfo{title}{Strong coupling corrections to the
  {G}inzburg-{L}andau theory of superfluid $^{3}${H}e}.
\newblock \emph{\bibinfo{journal}{Phys. Rev. B}} \textbf{\bibinfo{volume}{75}},
  \bibinfo{pages}{174503} (\bibinfo{year}{2007}).

\bibitem{Bun.93}
\bibinfo{author}{Bunkov, Y.~M.} \& \bibinfo{author}{Volovik, G.~E.}
\newblock \bibinfo{title}{On the possibility of the homogeneously precessing
  domain in bulk $^{3}${H}e-{A}}.
\newblock \emph{\bibinfo{journal}{Europhys. Lett.}}
  \textbf{\bibinfo{volume}{21}}, \bibinfo{pages}{837--843}
  (\bibinfo{year}{1993}).

\end{thebibliography}
\end{document}


\author{J. Pollanen}
\author{J.I.A. Li}
\author{C.A. Collett}
\author{W.J. Gannon}
\author{W.P. Halperin}
\email[Correspondence should be addressed to W.P.H.:~]{w-halperin@northwestern.edu}
\author{J.A. Sauls}
\affiliation{Department of Physics and Astronomy, Northwestern University, Evanston, Illinois 60208, USA}
\date{\today}

\title{Supplementary Information:\\
``New Chiral Phases of Superfluid \he\ Stabilized by Anisotropic Silica Aerogel''}
\maketitle

\section{Ginzburg-Landau theory}

The Ginzburg-Landau (GL) free-energy functional for $p$-wave, spin-triplet pairing of $^3$He quasiparticles
in a homogeneous isotropic medium is defined in Ref.~\cite{Thu.98} in terms of the $3\times 3$ matrix
order parameter, $A_{\delta i}$, which transforms as a vector under spin rotations with respect to index
$\delta$, separately as a vector under orbital rotations with respect to index $i$,
\begin{eqnarray}\label{eq:SI1.0}
\Delta\Omega_{\tiny{\mathsf{GL}}}[A] 
= \int_{\mathsf{V}}\,d^3r\,\Big\{
&&
\balpha(T) \Tr{AA^{\dag}} 
+\bbeta_1 \mid\Tr{AA^{\text{tr}}}\mid^2
+\bbeta_2 (\Tr{AA^{\dag}})^2
\nonumber\\
&&
+\bbeta_3 \Tr{AA^{\text{tr}}(AA^{\text{tr}})^*}
+\bbeta_4 \Tr{(AA^{\dag})^2} 
+\bbeta_5 \Tr{AA^{\dag}(AA^{\dag})^{\ast}}
\Big\}
\,.
\end{eqnarray}
The material coefficient, $\balpha(T)\simeq\balpha'(T-T_{ca})$, depends on the single spin density of states at the
Fermi surface, $N_{F}$, the pairing interaction, and the effects of scattering by the random impurity
potential, all of which determine the transition temperature, $T_{ca}$.  The prime denotes differentiation with respect to temperature.  The fourth-order material
parameters, $\{\bbeta_i\,|\, i=1\ldots 5\}$, which are also sensitive to impurity scattering, determine
condensation energy \underline{and} the structure and residual symmetry of the superfluid condensate. 

The homogeneous isotropic scattering model (HISM) of Thuneberg \emph{et al.} \cite{Thu.98} was extended
by Sauls and Sharma in Ref. \cite{Sau.03} to incorporate the effect of correlations between
silica strands and clusters. In this model, the aerogel correlation length, $\xi_a$, limits the pairing
transition for small superfluid coherence lengths, $\hbar v_F/2\pi k_B T_{ca} < \xi_a$ (high
pressures), while the mean free path, $\lambda$, determines the transition when the superfluid
coherence length is larger than the aerogel correlation length. This condition is achieved at low
pressures, and always in the limit $T_{ca}\rightarrow 0$. Indeed this limit can be used to determine
$\lambda$ from the critical pressure at which $T_{ca}(P_{c})=0$. The two pair-breaking length scales
define an effective pair-breaking parameter, $\tilde{x}=x/(1+\zeta^{2}_{a}/x)$,
$\zeta_{a}=\xi_{a}/\lambda$ where $x=\hbar v_{F}/2\pi k_{B}T\lambda$ is the depairing parameter in the
HISM \footnote{Our definition of $x$ is a factor of 2 larger than that in Ref.~\cite{Thu.98}.} and
$v_{F}$ is the pressure dependent Fermi velocity of \he\ \cite{Hal.90}. The superfluid transition
temperature in aerogel, $T_{ca}$, is determined by the condition $\balpha(T_{ca})=0$ in the following
equation,
\begin{equation}
			\bar{\alpha}(T)=\frac{1}{3}N_{F}\left[\ln(T/T_{c})-2\sum_{i=0}^\infty 
			\left(\frac{1}{2n+1+\tilde{x}}-\frac{1}{2n+1}\right)\right],
			\label{eq:SI1.1}
\end{equation}
where $T_{c}$ is the transition temperature of pure \he. In the limit $\tilde{x}=0$, Eq.~SI1 reduces to the case for pure \he, namely
$\alpha=(1/3)N_{F}\ln(T/T_{c})$.

The correlation length is a pressure independent length scale of the aerogel microstructure. We have used
Eq.~\ref{eq:SI1.1} to fit our measurements of $T_{ca}(P)$ in the stretched aerogel (Fig.~1c) with the
constraint that $\xi_{a}$ be pressure independent. From this fit we find $\lambda=113$ nm and
$\xi_{a}=39$ nm. In Fig.~SI1 we present the standard deviation of $\xi_{a}(P)$ as a function of
$\lambda$.
\begin{figure}
\centerline{\includegraphics[height=0.25\textheight]{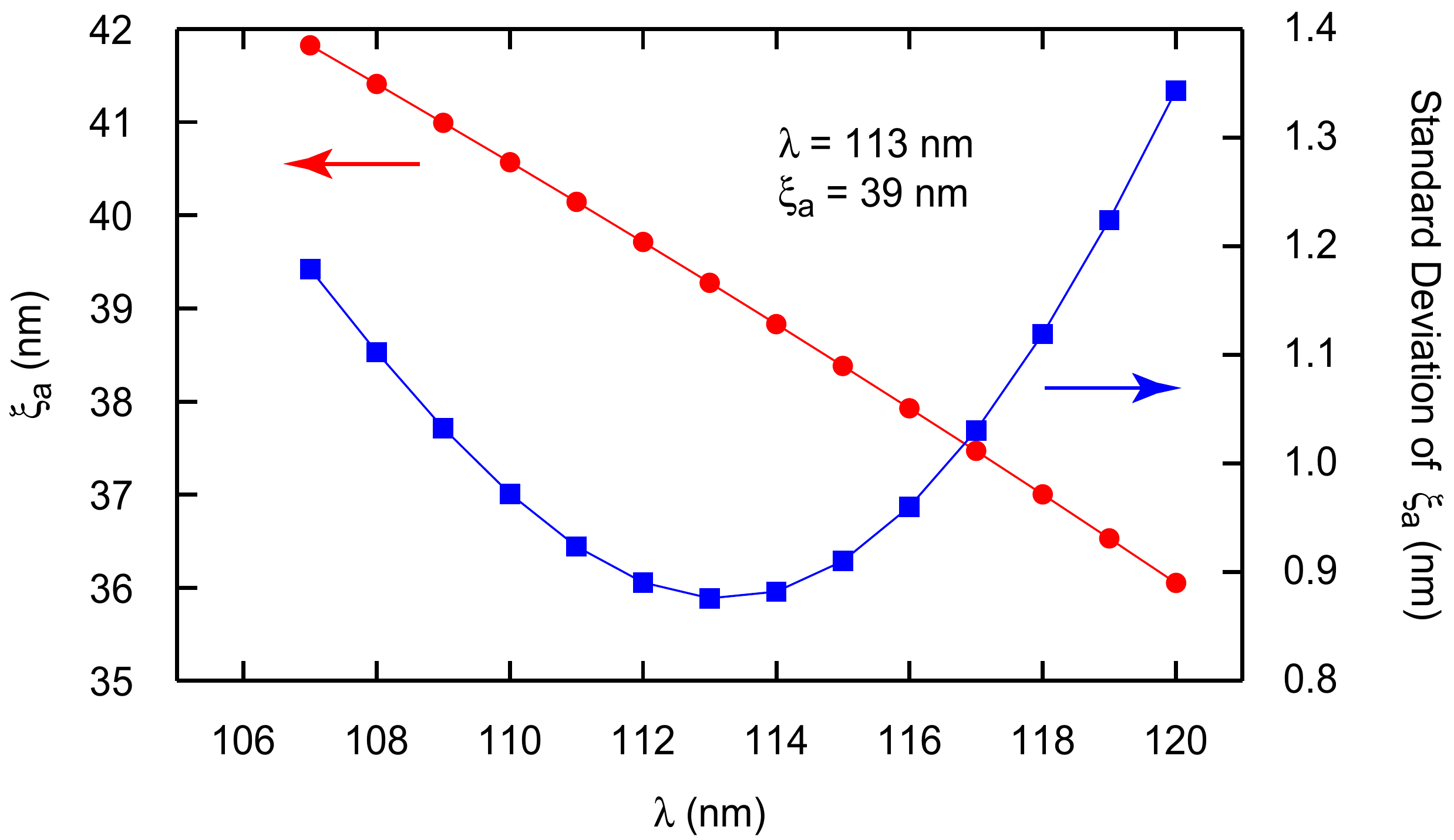}}
\caption{\label{figSI1.1}{\bf{Determination of Scattering Parameters}} Aerogel correlation length (red
circles) and standard deviation of $\xi_{a}$ as a function of $\lambda$ (blue squares). The minimum in
the standard deviation corresponds to $\lambda=113$ nm and $\xi_{a}=39$ nm.}
\end{figure}
Using these same values of $\lambda$ and $\xi_a$ we have calculated the square of the axial state order
parameter amplitude in aerogel from the GL theory,
\begin{equation}
\Delta^{2}_{Aa0}=\Delta^{2}_{A0} \hspace{1mm} \frac{T_{ca}}{T_{c}} 
	\hspace{1mm}\frac{\bar{\alpha}'(T_{ca})}{\alpha'(T_{c})} \hspace{1mm} \frac{\beta_{A}}{\bbeta_{A}}
\label{eq:SI1.2}
\end{equation}
In Eq.~SI2 $\Delta^{2}_{A0}$ is the value
for the pure A-phase \cite{Sch.93, Ran.96}, and $\beta_{A}$ ($\bbeta_{A}$) is the combination of
$\beta$-parameters appropriate to pure (aerogel) \he-A, \emph{i.e.}
$\beta_{A}=\beta_{2}+\beta_{4}+\beta_{5}$ ($\bbeta_{A}=\bbeta_{2}+\bbeta_{4}+\bbeta_{5}$). The pure
$\beta$-parameters were taken from Ref.~\cite{Cho.07}. For calculating the aerogel $\bbeta$-parameters,
we have extended the results of Thuneberg \et\ \cite{Thu.98} for the HISM, to include the combined
effects of the aerogel correlation length and mean-free path using the correlated scattering
model of Ref.~\cite{Sau.03}.

\section{NMR Line Shape Analysis}

We have performed an analysis of the NMR line shape to obtain the distribution of the chiral axis,
$\vec{\ell}$, in our stretched sample. A distribution in the direction of $\vec{\ell}$ relative to the
external magnetic field, $\vec{H}$, will lead to a distribution in the intrinsic superfluid dipole field
and hence a distribution of frequency shifts in the NMR spectrum. If $\theta=\acos({\vec{\ell} \cdot
\vec{H}})$ is the angle between the chiral axis and the magnetic field, then for small tip angle,
$\beta$, the frequency shift in the A-phase reduces to \cite{Bun.93},
\begin{equation}
		\Delta\omega_{A}(\theta)=-\frac{\Omega^{2}_{A}}{2\omega_{L}}\cos(2\theta).
		\label{eq:SI2.1}
\end{equation}

\noindent
For example, if the chiral axis orientation is unique and the normal state lineshape is given by
$F_{n}(\omega-\omega_{L})$ then the lineshape in the superfluid, $F_{s}$, is unchanged except for a
constant shift, \emph{i.e.} $F_{s}=F_{n}(\omega-\omega_{L}-\Delta\omega_{A})$ where $\Delta\omega_{A}$ is
given by Eq.~SI3. However, for an angular distribution of $\vec{\ell}$, $P(\theta)$, there is a
distribution of frequency shifts, $P(\omega)$, determined by Eq.~SI3. This results in a more complex NMR
lineshape that is given by the convolution product of the normal state line and $P(\omega)$,
	\begin{equation}
	F_{s}(\omega)=\int \! P(\omega^{'})F_{n}\left(\omega-\omega_{L}-\omega^{'}\right) \,d\omega^{'}.
	\label{eqSI2.2}
	\end{equation}
Since the normal state and superfluid spectra are obtained experimentally, $P(\theta)$ can be determined
by fitting the spectrum in the superfluid phase with Eq.~SI4.

In general an inversion problem, like this one, requires constraints which make it difficult to determine
the uniqueness of the distribution. We choose the model that the $\vec{\ell}$-distribution can be
represented by a sum of gaussian functions with adjustable position, $\theta_{i}$, width, $w_{i}$, and
relative weight, $A_{i}$,
\begin{equation}
     P\left(\theta \right) = \sum_{i}A_{i}e^{\left(\frac{\theta-\theta_i}{w_i}\right)^2},
     \label{eq6.7}
\end{equation}
and $\int \! P\left ( \theta\right ) \,d\theta = 1$. In Fig.~4a we present NMR spectra (bold green
curves) and corresponding fits (dashed black curves) for temperatures above (left panel) and below (right
panel) the disorder transition $T_{d}$ with the results for $P(\theta)$ shown in Fig.~4b. For
$T_{d}<T<T_{ca}$ we find that the spectrum is well-represented with a distribution centered at $\theta =
90^\circ$, corresponding to the dipole-locked configuration, with a spread of $13.0^{\circ} \pm 2.3^{\circ}$. For $T<T_{d}$, we find that the
$\vec{\ell}$-distribution is composed of domains as described in the text. Approximately $1/3$ of the
sample remains with $\vec{\ell}$ perpendicular to $\vec{H}$ with a spread $\lesssim 4^{\circ}$
as shown in Fig.~SI2. where we present the squared residual of the fit as a function of the width of the
dipole-locked peak while keeping the other parameters fixed. The remaining $\sim2/3$ of the superfluid
has $\theta=44^{\circ} \pm 0.5^{\circ}$ with a width of $4.9^{\circ}\pm 0.6^{\circ}$ as shown in
Fig.~SI3.
\noindent
\begin{figure}
\centerline{\includegraphics[height=0.225\textheight]{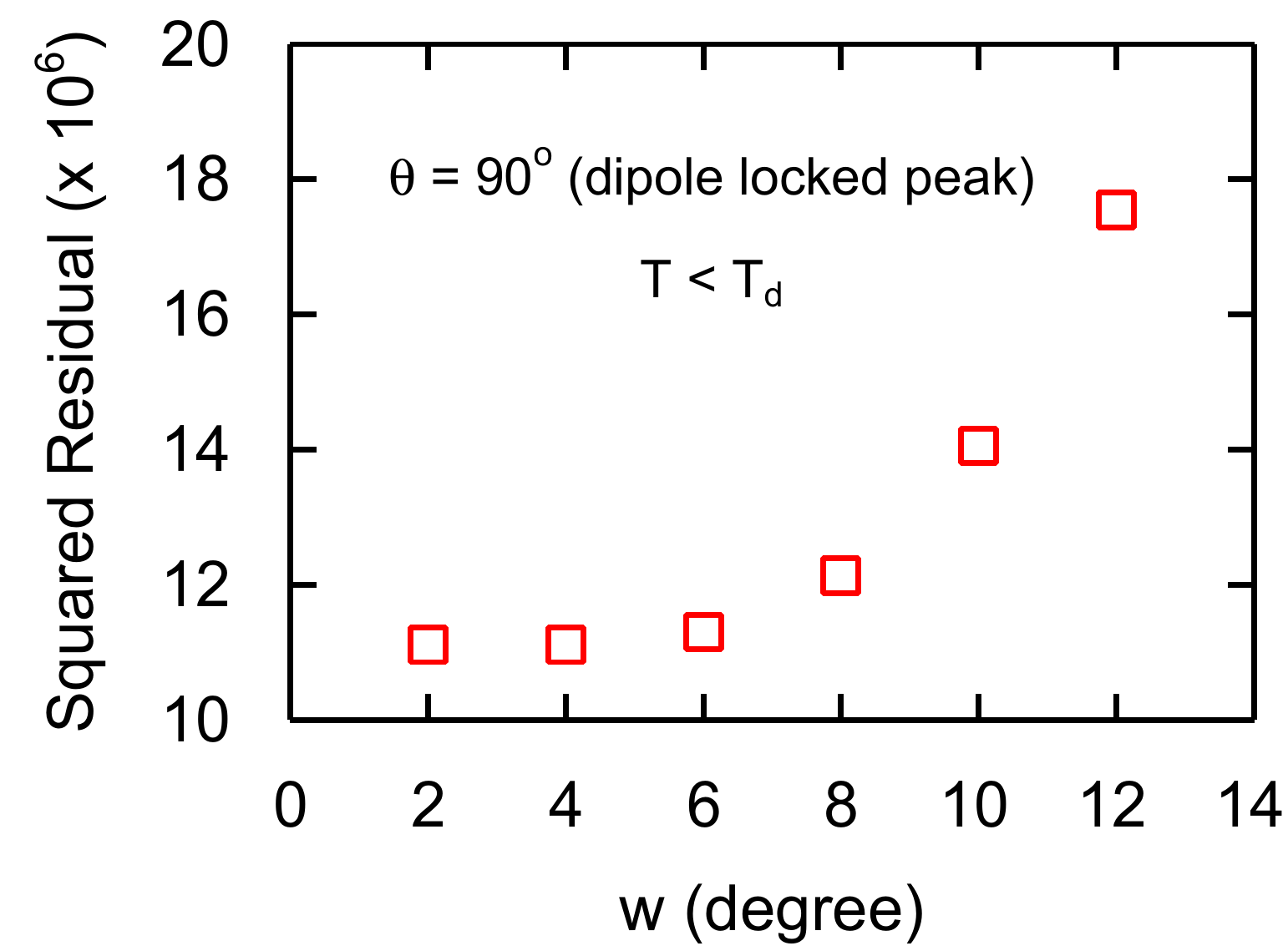}}
\caption{\label{figSI2.1}Squared residual versus the width of the dipole-locked peak for $T<T_{d}$.}
\end{figure}
\noindent
This distribution fits our spectrum well for $T<T_{d}$. To check its uniqueness we have attempted to add
an additional peak with variable relative weight, $w_{v}$, at the position $\theta=23^{\circ}$, with a
width of $\theta=6^{\circ}$ and we find that the best fit to the spectrum corresponds to $w_{v}=0$ as
shown in Fig.~SI4a. To emphasize this point, in Fig.~SI4b we show the lineshape that results for the case
$w_{v} = 15\%$ that clearly adds substantial weight in the spectrum which is inconsistent with our
experiment. We infer, therefore, that the disordered state has an $\vec{\ell}$-distribution which is
non-monotonic with three principal components as shown in Fig. 4b.
\noindent
\begin{figure}
\centerline{\includegraphics[height=0.25\textheight]{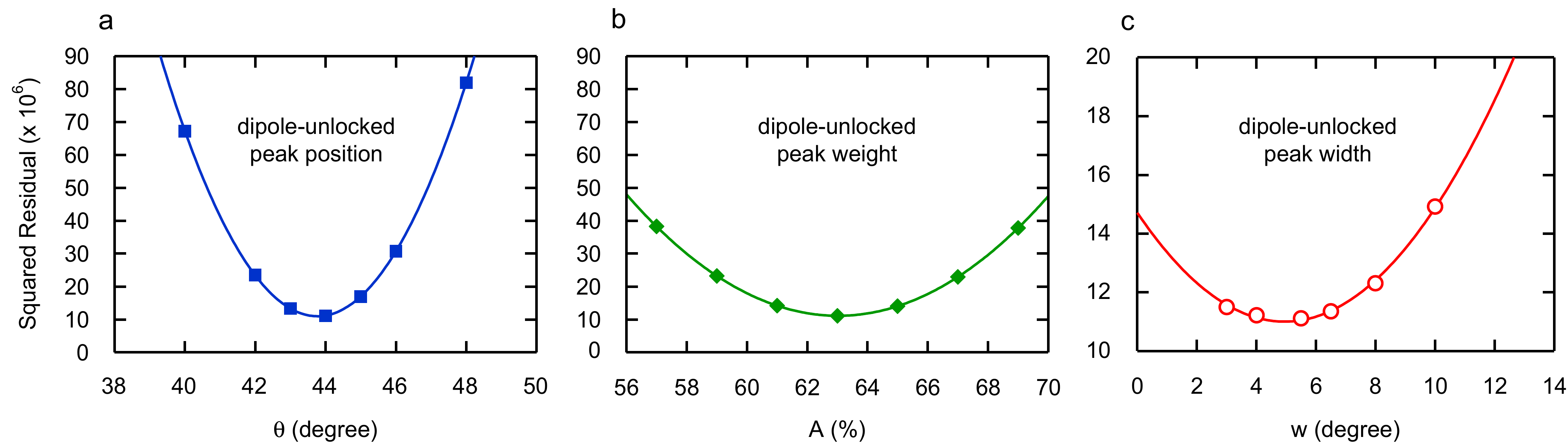}}
\caption{\label{figSI2.2}Squared residual versus the a) position, b) percent weight, and c) width of the
dipole-unlocked peak for $T<T_{d}$.}
\end{figure}
\noindent
\begin{figure}
\centerline{\includegraphics[height=0.25\textheight]{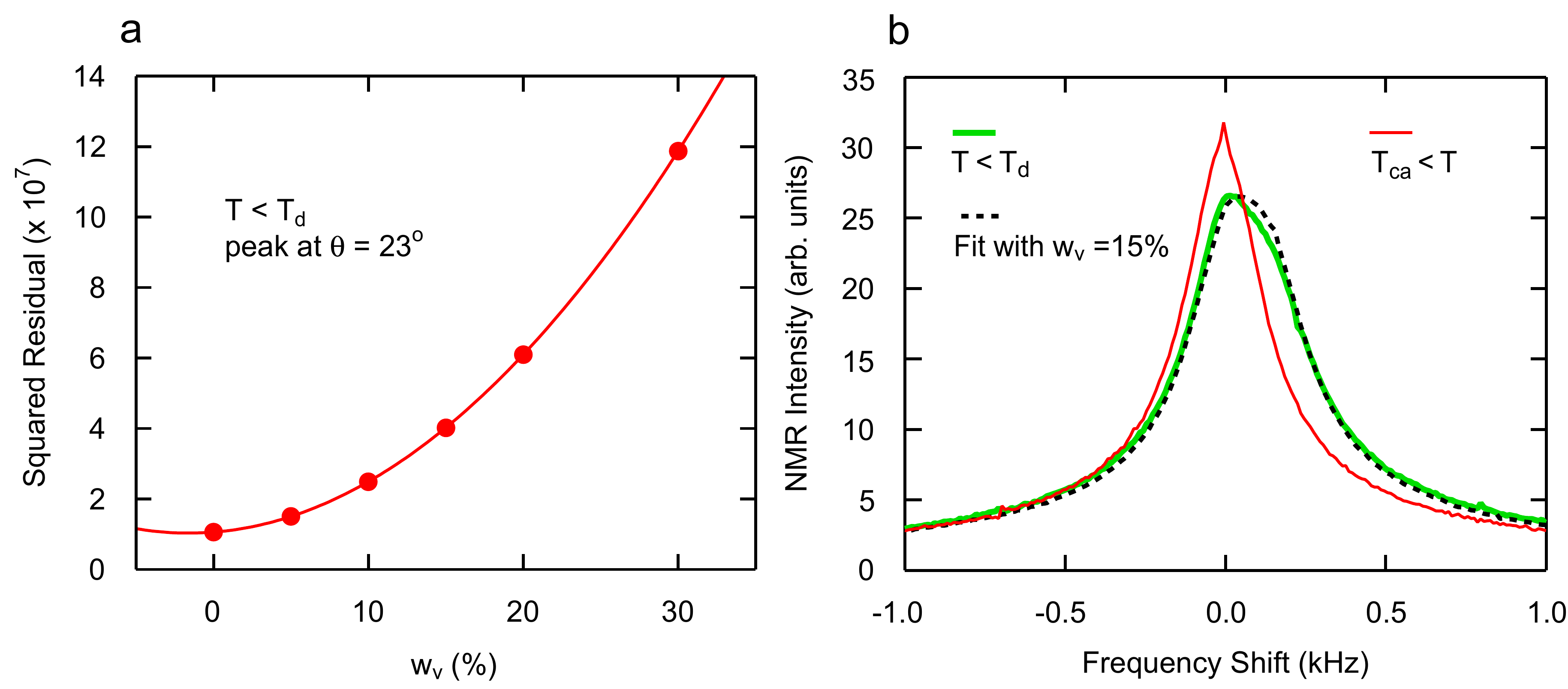}}
\caption{\label{figSI2.3}{\bf{Attempt to add an additional peak}} a) Squared residual versus the weight of an
additional peak at $\theta=23^{\circ}$. b) Fit (dashed black curve) to the NMR spectrum below $T_{d}$
(bold green curve) produced with $w_{v}=15\%$. The normal state line is presented as the fine red curve
and the fit is performed using Eq.~SI4. Significant discrepancy between the best fit and the NMR spectrum
indicates that extra weight in the $\vec{\ell}$-distribution between $44^{\circ}$ and $90^{\circ}$ can be
ruled out.}
\end{figure}
 
\bibliography{NPPollRef}